\documentclass[twocolumn,superscriptaddress,letter,prb]{revtex4}%
\usepackage{amssymb}
\usepackage{color}
\usepackage{graphicx}
\usepackage{dcolumn}
\usepackage{bm}
\usepackage[header,title,page,titletoc]{appendix}
\usepackage{amsmath}
\usepackage{amsfonts}%
\providecommand{\U}[1]{\protect \rule{.1in}{.1in}}

\begin{document}
\title{Anomalous Spin Dynamics of Hubbard Model on Honeycomb Lattices}
\author{Gao-Yong Sun}
\affiliation{Department of Physics, Beijing Normal University, Beijing 100875, China}
\author{Su-Peng Kou}
\thanks{Corresponding author}
\email{spkou@bnu.edu.cn}
\affiliation{Department of Physics, Beijing Normal University, Beijing 100875, China}

\begin{abstract}
In this paper, the honeycomb Hubbard model in optical lattices is investigated
using O(3) non-linear $\sigma$ model. A possible quantum non-magnetic
insulator in a narrow parameter region is found near the metal-insulator
transition. We study the corresponding dynamics of magnetic properties, and
find that the narrow region could be widened by hole doping.


\end{abstract}
\maketitle

\section{Introduction}

Recently, using ultracold atoms to form Bose-Einstein Condensates (BEC) or
Fermi degenerate gases for precise measurements and\ simulations of quantum
many-body systems, is quite impressive and has become a rapidly-developing
field\cite{Bloch,review}. Experimental realizations of quantum many-body
systems in optical lattices have made a chance to simulate strongly correlated
systems\cite{Bloch}. People have successfully observed the Mott
insulator--superfluid transition both in bosonic\cite{Greiner} and fermionic
atoms (e.g. $^{6}$Li or $^{40}$K, etc.)\cite{Esslinger,Bloch2}. In particular,
the two-dimensional (fermionic) Hubbard model is one of the most interesting
issues depicting the nature of high-temperature superconductivity. It is still
a big challenge to clearly understand the physics of repulsive Hubbard model
on two dimensional (2D) lattices. Thus people try to simulate the
Fermi-Hubbard model using a two-component mixture of repulsively interacting
fermions\cite{Iskin,Andersen,Snoek,Gottwald}.

In this paper, we focus on the two dimensional Hubbard model in honeycomb
lattices (2D honeycomb Hubbard model). In Refs.\cite{duan,Zhu,Lee}, it is
proposed that the 2D honeycomb optical lattice can be realized in the cold
atoms with three detuned standing-wave lasers, of which the optical potential
is given by
\begin{equation}
V(x,y)=\sum_{j=1,2,3}V\sin^{2}[k_{L}(x\cos \theta_{j}+y\sin \theta_{j})+\pi/2]
\end{equation}
where $\theta_{1}=\pi/3,$ $\theta_{2}=2\pi/3,$ $\theta_{3}=0$, and $k_{L}$ is
the optical wave vector. When two-component fermions with repulsive
interaction are put into the 2D honeycomb optical lattice, one can get an
effective honeycomb Hubbard model. It is easy to change the potential barrier
$V$ by varying the laser intensities to tune the Hamiltonian parameters
including the hopping strength $t$ ($t$-term) and the particle interaction $U$
($U$-term). This lays the foundation for our later discussions and calculations.

\begin{figure}[ptb]
\includegraphics[width=0.45\textwidth]{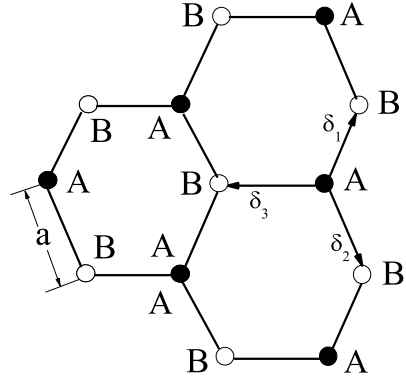}\caption{Illustration of a
honeycomb\textit{\ }lattice with A and B sublattices. $a$ is the length of the
hexagon side and chosen to be unit.}%
\end{figure}

The simplest approach to study the honeycomb Hubbard model is the Hartree-Fock
(HF) mean field method, from which people obtain a semi-metal-insulator (MI)
transition at a critical value $(U/t)_{c}$ between a semi-metal (SM) and an
antiferromagnetic (AF) insulator. In the weak interaction region
$U/t<(U/t)_{c}$, the ground state is a semi-metal (SM) with nodal
fermi-points. In the strong interaction region $U/t>(U/t)_{c}$, the ground
state becomes an insulator with massive fermionic excitations. However, the HF
theory does not keep the spin rotation symmetry by fixing the spins along
$\mathbf{\hat{z}}$-axis. So the results are not reliable. Though the
semi-metal-insulator transition of the honeycomb Hubbard model has been
studied by different approaches,\cite{Hermele,Sorella,Martelo,Paiva} the
results are not consistent with each other.

In this paper, we will investigate the two dimensional honeycomb Hubbard model
by an approach proposed in Refs.\cite{wen,dup,dupuis,sun,Schulz,weng} that
keeps spin rotation symmetry. By it, we find anomalous spin dynamics not far
from the critical point of MI transition : there\ may exist a narrow
non-magnetic insulator. The narrow non-magnetic insulator will be in favor of hole-doping.

The paper is organized as follows. In Sec. II, the semi-metal-insulator
transition is studied by HF mean field approach. In Sec. III, an effective
O(3) non-linear $\sigma$ model ($\mathrm{NL}\sigma \mathrm{M}$) is obtained to
investigate properties of the honeycomb-Hubbard model. In Sec. IV, a global
phase diagram is given and magnetic properties of the insulator state is
studied based on the $\mathrm{NL}\sigma \mathrm{M.}$ In Sev. V, we discuss how
to observe the anomalous spin dynamics in optical lattice of cold atoms. In
Sec. VI, we show the doping effect on the magnetic properties of the ground
state. Finally, the conclusions are given in Sec. VII.

\section{Metal--insulator transition}

As a starting point, the Hamiltonian of the Hubbard model on honeycomb lattice
is given by
\begin{equation}
\mathcal{H}=-t\sum \limits_{\langle i,j\rangle}\left(  \hat{c}_{i}^{\dagger
}\hat{c}_{j}+h.c.\right)  +U\sum_{i}\hat{n}_{i\uparrow}\hat{n}_{i\downarrow
}-\mu \sum \limits_{i}\hat{c}_{i}^{\dagger}\hat{c}_{i}. \label{mo}%
\end{equation}
Here $\hat{c}_{i}=(\hat{c}_{i\uparrow},\hat{c}_{i\downarrow})^{T}$ and
$\hat{c}_{i,\sigma}^{\dagger},$ $\hat{c}_{j,\sigma}$ are electronic creation
and annihilation operators. $t$ is the hopping integral. $U$ is the on-site
Coulomb repulsion. $\sigma$ are the spin-indices representing spin-up($\sigma
=\uparrow$) and spin-down($\sigma=\downarrow$) for electrons. $\mu$ is the
chemical potential which is $\frac{U}{2}$ in half-filling. $\langle
i,j\rangle$ denotes two sites on a nearest-neighbor link. $\hat{n}_{i\uparrow
}$ and $\hat{n}_{i\downarrow}$ are the number operators of electrons with
up-spin and down spin respectively.

Because the honeycomb lattice is a bipartite lattice (See Fig.1), we divide
the system into two sublattices, A and B. Using the Fourier transformations,
the electronic annihilation operators on two sublattices are written into%
\begin{align}
\hat{c}_{i\in A,\sigma}  &  =\frac{1}{\sqrt{N_{s}}}\sum \limits_{\mathbf{k}%
}e^{-i\mathbf{k\cdot R}_{i}}\hat{a}_{\mathbf{k}\sigma},\\
\hat{c}_{i\in B,\sigma}  &  =\frac{1}{\sqrt{N_{s}}}\sum \limits_{\mathbf{k}%
}e^{-i\mathbf{k\cdot R}_{i}}\hat{b}_{\mathbf{k}\sigma},
\end{align}
where $N_{s}$ denoting the number of unit cells. For free fermions, the
Hamiltonian could be transformed in the momentum space as%
\begin{equation}
\mathcal{H}=\sum \limits_{\mathbf{k,}\sigma}\left(
\begin{array}
[c]{cc}%
\hat{a}_{\mathbf{k}\sigma}^{\dagger} & \hat{b}_{\mathbf{k}\sigma}^{\dagger}%
\end{array}
\right)  \left(
\begin{array}
[c]{cc}%
0 & \xi_{\mathbf{k}}\\
\xi_{\mathbf{k}}^{\ast} & 0
\end{array}
\right)  \left(
\begin{array}
[c]{c}%
\hat{a}_{\mathbf{k}\sigma}\\
\hat{b}_{\mathbf{k}\sigma}%
\end{array}
\right)
\end{equation}
where the energy of free fermions is%
\begin{align}
\left \vert \xi_{\mathbf{k}}\right \vert  &  =\left \vert -t\sum \limits_{\delta
}e^{i\mathbf{k\cdot \delta}}\right \vert \label{kesi}\\
&  =t\sqrt{3+2\cos \left(  \sqrt{3}k_{y}\right)  +4\cos \left(  3k_{x}/2\right)
\cos \left(  \sqrt{3}k_{y}/2\right)  }.\nonumber
\end{align}
Here the nearest neighbors of a electron in A lattice are defined as
\begin{equation}
\delta_{1}=\frac{a}{2}\left(  1,\sqrt{3}\right)  ,\text{ }\delta_{2}=\frac
{a}{2}\left(  1,-\sqrt{3}\right)  ,\text{ }\delta_{3}=\left(  -a,0\right)
\end{equation}
where $a$ is the length of the hexagon side and chosen to be unit. The
spectrum for free electrons is then obtained as $E_{\mathbf{k}}=\pm \left \vert
\xi_{\mathbf{k}}\right \vert .$

Next we use the path-integral formulation of electrons with spin rotation
symmetry to study the on-site repulsive interaction in Hubbard
model.\cite{wen,dup,dupuis,sun,Schulz,weng} The interaction term can be
handled by using a SU(2) invariant Hubbard-Stratonovich (HS) decomposition in
the arbitrary on-site unit vector $\mathbf{\Omega}_{i}$%
\begin{equation}
\hat{n}_{i\uparrow}\hat{n}_{i\downarrow}=\frac{\left(  \hat{c}_{i}^{\dagger
}\hat{c}_{i}\right)  ^{2}}{4}-\frac{1}{4}[\mathbf{\Omega}_{i}\mathbf{\cdot
}\hat{c}_{i}^{\dag}\mathbf{\sigma}\hat{c}_{i}]^{2}%
\end{equation}
where $\mathbf{\sigma=}\left(  \mathbf{\sigma}_{x},\mathbf{\sigma}%
_{y},\mathbf{\sigma}_{z}\right)  $ is the Pauli matrix. Then the HS
transformation for the interaction term is%

\begin{align}
e^{U\sum_{i}\hat{n}_{i\uparrow}\hat{n}_{i\downarrow}}  &  =\int \prod_{i}%
\frac{d\Delta_{c}d\Delta_{i}d^{2}\mathbf{\Omega}_{i}}{4\pi^{2}U}\exp(\sum
_{i}[\frac{1}{U}\left(  \Delta_{c}^{2}+\Delta_{i}^{2}\right) \nonumber \\
&  +i\Delta_{c}\hat{c}_{i}^{\dag}\hat{c}_{i}-\Delta_{i}\hat{c}_{i}^{\dag
}\mathbf{\Omega}_{i}\mathbf{\cdot \sigma}\hat{c}_{i}]).
\end{align}
Here $\Delta_{c}$ and $\Delta_{i}$ are the auxiliary fields. By replacing
electronic operators $\hat{c}_{i,\sigma}^{\dagger}$ and $\hat{c}_{j,\sigma}$
to Grassmann variables $c_{i,\sigma}^{\ast}$ and $c_{j,\sigma}$, the effective
Lagrangian in terms of Grassmann variables $c_{i,\sigma}^{\ast}$ and
$c_{i,\sigma}$ is then obtained as
\begin{align}
\mathcal{L}_{\mathrm{eff}}  &  =\sum_{i,\sigma}c_{i,\sigma}^{\ast}%
\partial_{\tau}c_{i,\sigma}-\sum \limits_{\left \langle ij\right \rangle
}(t_{i,j}c_{i}^{\ast}c_{j}+h.c.)-\sum_{i}\Delta_{i}c_{i}^{\ast}\mathbf{\Omega
}_{i}\mathbf{\cdot \sigma}c_{i}\nonumber \\
&  +\sum_{i}\left[  \frac{1}{U}\left(  \Delta_{c}^{2}+\Delta_{i}^{2}\right)
+\left(  i\Delta_{c}-\mu \right)  c_{i}^{\ast}c_{i}\right]  .
\end{align}

The ground state of the honeycomb Hubbard model is known to be long-range AF
order in the large $U$ limit. Such \textbf{an} order can be described by a
simple saddle-point Lagrangian by fixing the direction vector field
$\mathbf{\Omega}_{i}$ to $\mathbf{\hat{z}}$-axis $\mathbf{\Omega}_{i}%
=(-1)^{i}\mathbf{\hat{z}}$ and choosing the amplitude as
\begin{align}
i\Delta_{c}  &  =\frac{U}{2}\left \langle c_{i}^{\dagger}c_{i}\right \rangle
=\frac{Un}{2}\label{n}\\
\Delta_{i}  &  =\frac{U}{2}\left \langle c_{i}^{\dagger}\mathbf{\sigma}%
_{z}c_{i}\right \rangle =\left(  -1\right)  ^{i}\frac{UM}{2}=\left(  -1\right)
^{i}\Delta \label{sta}%
\end{align}
where $n$ is the average on-site electron density and cancels with the
chemical potential $\mu$ in half-filled ($n=1$) case. $M$ is the staggered
magnetization and $\Delta=\frac{UM}{2}$ is the energy band gap. The effective
Lagrangian turns into%

\begin{equation}
\mathcal{L}_{\mathrm{eff}}=\sum_{i,\sigma}c_{i,\sigma}^{\ast}\partial_{\tau
}c_{i,\sigma}-\sum \limits_{\left \langle ij\right \rangle }(t_{i,j}c_{i}^{\ast
}c_{j}+h.c.)-\sum_{i}\left(  -1\right)  ^{i}\Delta c_{i}^{\ast}\mathbf{\sigma
}_{z}c_{i}. \label{mean}%
\end{equation}
One may obtain the spectrum of the electrons as%
\begin{equation}
E_{\mathbf{k}}=\pm \sqrt{\left \vert \xi_{\mathbf{k}}\right \vert ^{2}+\Delta
^{2}}%
\end{equation}
Finally we derive the self-consistency equation for $M$ by minimizing the free
energy at temperature $T$ in the Brillouin zone as
\begin{equation}
1=\frac{1}{N}\sum \limits_{\mathbf{k}}\frac{U}{2E_{\mathbf{k}}}\tanh \left(
\beta E_{\mathbf{k}}/2\right)  \label{self}%
\end{equation}
Here $\beta=\frac{1}{k_{B}T}$. $N$ is the number of the sites.
\begin{figure}[ptb]
\includegraphics[width=0.45\textwidth]{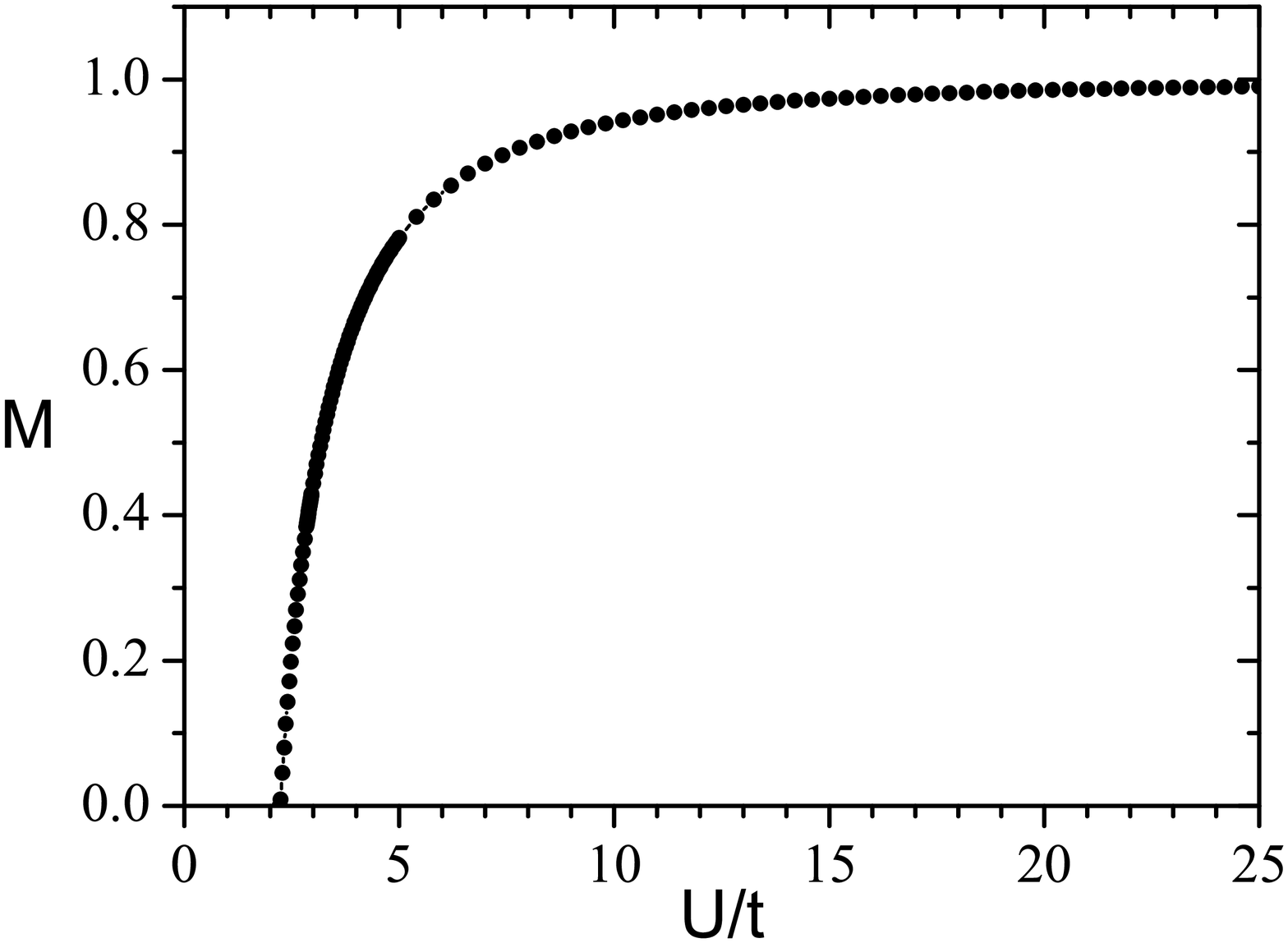}\caption{Staggered
magnetization $M$ of half-filled honeycomb lattice Hubbard model at $T=0$.}%
\end{figure}

From Fig.2, one could find that MI transition occurs at a critical value about
$U/t\simeq2.23$ at zero temperature\cite{Sorella,Paiva,Martelo,Peres}. In the
weakly coupling limit $\left(  U/t<2.23\right)  $, the ground state is a
semi-metal (SM) with \emph{nodal} fermi-points. In the strong coupling region
$\left(  U/t>2.23\right)  $, due to $M\neq0$, the ground state becomes an
insulator with massive fermionic excitations. However, the non-zero value of
$M$ only means the existence of effective spin moments rather than a long
range AF order since this result is obtained in a mean field level by fixing
the spins along $\mathbf{\hat{z}}$-axis. Thus one needs to examine stability
of magnetic order against quantum fluctuations of effective spin moments by
keeping spin rotation symmetry.

\section{Effective nonlinear $\sigma$ model in the insulator state}

In this section, we will derive an effective $\mathrm{NL}\sigma \mathrm{M}$ of
spin fluctuations with spin rotation symmetry in the honeycomb Hubbard model
beyond above mean field theory.

To describe the spin fluctuations, we use the Haldane's mapping:
\begin{equation}
\mathbf{\Omega}_{i}=(-1)^{i}\mathbf{n}_{i}\sqrt{1-\mathbf{L}_{i}^{2}%
}+\mathbf{L}_{i}%
\end{equation}
where $\mathbf{n}_{i}$ is the Neel vector that corresponds to the
long-wavelength part of $\mathbf{\Omega}_{i}$ with a restriction
$\mathbf{n}_{i}^{2}=1.$ $\mathbf{L}_{i}$ is the transverse canting field that
corresponds to the short-wavelength parts of $\mathbf{\Omega}_{i}$ with a
restriction $\mathbf{L}_{i}\cdot \mathbf{n}_{i}=0$%
\cite{dupuis,Haldane,Auerbach}. We then rotate $\mathbf{\Omega}_{i}$ to
$\mathbf{\hat{z}}$-axis for the spin indexes of the electrons at
$i$-site:\cite{wen,dup,dupuis,sun,Schulz,weng}%
\begin{align}
\psi_{i}  &  =U_{i}^{\dagger}c_{i}\nonumber \\
U_{i}^{\dagger}\mathbf{n}_{i}\cdot \mathbf{\sigma}U_{i}  &  =\mathbf{\sigma
}_{z}\nonumber \\
U_{i}^{\dagger}\mathbf{L}_{i}\cdot \mathbf{\sigma}U_{i}  &  =\mathbf{l}%
_{i}\cdot \mathbf{\sigma}%
\end{align}
where $U_{i}\in$\textrm{SU(2)/U(1)}. One then can derive the following
effective Lagrangian after such spin transformation:
\begin{align}
\mathcal{L}_{\mathrm{eff}}  &  =\sum \limits_{i}\psi_{i}^{\ast}\partial_{\tau
}\psi_{i}+\sum \limits_{i}\psi_{i}^{\ast}a_{0}\left(  i\right)  \psi
_{i}\nonumber \\
&  -\sum \limits_{<ij>}(t_{i,j}\psi_{i}^{\ast}e^{ia_{ij}}\psi_{j}%
+h.c.)\nonumber \\
&  -\Delta \sum \limits_{i}\psi_{i}^{\ast}\left[  (-1)^{i}\mathbf{\sigma}%
_{z}\sqrt{1-\mathbf{l}_{i}^{2}}+\mathbf{l}_{i}\cdot \mathbf{\sigma}\right]
\psi_{i} \label{lag}%
\end{align}
where the auxiliary gauge fields $a_{ij}=a_{ij,1}\sigma_{x}+a_{ij,2}\sigma
_{y}$ and $a_{0}\left(  i\right)  =a_{0,1}\left(  i\right)  \sigma_{x}%
+a_{0,2}\left(  i\right)  \sigma_{y}\ $are defined:
\begin{equation}
e^{ia_{ij}}=U_{i}^{\dagger}U_{j},\text{ }a_{0}\left(  i\right)  =U_{i}%
^{\dagger}\partial_{\tau}U_{i}. \label{ao}%
\end{equation}

In terms of the mean field result $M=\left(  -1\right)  ^{i}\langle \psi
_{i}^{\ast}\mathbf{\sigma}_{z}\psi_{i}\rangle$ as well as the approximations,
\begin{equation}
\sqrt{1-\mathbf{l}_{i}^{2}}\simeq1-\frac{\mathbf{l}_{i}^{2}}{2},\text{
}e^{ia_{ij}}\simeq1+ia_{ij},\nonumber
\end{equation}
we obtain the effective Hamiltonian as:
\begin{align}
\mathcal{L}_{\mathrm{eff}}  &  \simeq \sum \limits_{i}\psi_{i}^{\ast}%
\partial_{\tau}\psi_{i}+\sum \limits_{i}\psi_{i}^{\ast}[a_{0}\left(  i\right)
-\Delta \mathbf{\sigma \cdot l}_{i}]\psi_{i}\nonumber \\
&  -\sum \limits_{\left \langle ij\right \rangle }[t_{i,j}\psi_{i}^{\ast
}(1+ia_{ij})\psi_{j}+h.c.]\nonumber \\
&  -\Delta \sum \limits_{i}(-1)^{i}\psi_{i}^{\ast}\sigma_{z}\psi_{i}+\Delta
M\sum \limits_{i}\frac{\mathbf{l}_{i}^{2}}{2}.
\end{align}
By integrating out the fermion fields $\psi_{i}^{\ast}$ and $\psi_{i},$ the
effective action with the quadric terms of $[a_{0}\left(  i\right)
-\Delta \mathbf{\sigma \cdot l}_{i}]$ and $a_{ij}$ becomes
\begin{equation}
\mathcal{S}_{\mathrm{eff}}=\frac{1}{2}\int_{0}^{\beta}d\tau \sum_{i}%
[-4\varsigma(a_{0}\left(  i\right)  -\Delta \mathbf{\sigma \cdot l}_{i}%
)^{2}+4\rho_{s}a_{ij}^{2}+\frac{2\Delta^{2}}{U}\mathbf{l}_{i}^{2}]
\label{efff}%
\end{equation}
where $\rho_{s}$ and $\varsigma$ are two parameters.

To learn the properties of the low energy physics, we study the continuum
theory of the effective action in Eq.(\ref{efff}). In the continuum limit, we
denote $\mathbf{n}_{i}$, $\mathbf{l}_{i}$, $ia_{ij}\simeq U_{i}^{\dagger}%
U_{j}-1$ and $a_{0}\left(  i\right)  =U_{i}^{\dagger}\partial_{\tau}U_{i}$ by
$\mathbf{n}(x,y)$, $\mathbf{l}(x,y)$, $U^{\dagger}\partial_{x}U$ (or
$U^{\dagger}\partial_{y}U$) and $U^{\dagger}\partial_{\tau}U,$ respectively.
From the relations between $U^{\dagger}\partial_{\mu}U$ and $\partial_{\mu
}\mathbf{n,}$
\begin{align}
a_{\tau}^{2}  &  =a_{\tau,1}^{2}+a_{\tau,2}^{2}=-\frac{1}{4}(\partial_{\tau
}\mathbf{n})^{2},\text{ }\tau=0,\label{tao}\\
a_{\mu}^{2}  &  =a_{\mu,1}^{2}+a_{\mu,2}^{2}=\frac{1}{4}(\partial_{\mu
}\mathbf{n})^{2},\text{ }\mu=x,y,\label{miu}\\
\mathbf{a}_{0}\mathbf{\cdot l}  &  \mathbf{=}\mathbf{-}\frac{i}{2}\left(
\mathbf{n}\times \partial_{\tau}\mathbf{n}\right)  \cdot \mathbf{l,} \label{dot}%
\end{align}
the continuum formulation of the action in Eq.(\ref{efff}) turns into
\begin{align}
\mathcal{S}_{\mathrm{eff}}  &  =\frac{1}{2}\int_{0}^{\beta}d\tau \int
d^{2}\mathbf{r}[\varsigma(\partial_{\tau}\mathbf{n)}^{2}+\rho_{s}\left(
\mathbf{\bigtriangledown n}\right)  ^{2}\nonumber \\
&  -4i\Delta \varsigma \left(  \mathbf{n}\times \partial_{\tau}\mathbf{n}\right)
\cdot \mathbf{l}+(\frac{2\Delta^{2}}{U}-4\Delta^{2}\varsigma)\mathbf{l}^{2}]
\label{sigma}%
\end{align}
where the vector $\mathbf{a}_{0}$ is defined as $\mathbf{a}_{0}=\left(
a_{0,1},\text{ }a_{0,2},\text{ }0\right)  .$

Finally we integrate the transverse canting field $\mathbf{l}$ and obtain the
effective $\mathrm{NL}\sigma \mathrm{M}$ as
\begin{equation}
\mathcal{S}_{\mathrm{eff}}=\frac{1}{2g}\int_{0}^{\beta}d\tau \int d^{2}%
r[\frac{1}{c}\left(  \partial_{\tau}\mathbf{n}\right)  ^{2}+c\left(
\mathbf{\bigtriangledown n}\right)  ^{2}]\text{ } \label{non}%
\end{equation}
with a constraint $\mathbf{n}^{2}=1.$ The coupling constant $g$ and spin wave
velocity $c$ are defined as:
\begin{equation}
g=\frac{c}{\rho_{s}},\text{ }c^{2}=\frac{\rho_{s}}{\chi^{\perp}}\text{, }%
\chi^{\perp}=(\frac{1}{\zeta}-2U)^{-1}.
\end{equation}
Here $\rho_{s}$ is the spin stiffness,
\begin{equation}
\rho_{s}=\frac{1}{N}\sum \limits_{\mathbf{k}}\frac{\epsilon^{2}}{4(\left \vert
\xi_{\mathbf{k}}\right \vert ^{2}+\Delta^{2})^{\frac{3}{2}}},
\end{equation}
where the corresponding coefficient $\epsilon^{2}$ is
\begin{align}
\epsilon^{2}  &  =\frac{1}{4}t^{2}[6\Delta^{2}+27t^{2}+\left(  2\Delta
^{2}+27t^{2}\right)  \cos \left(  \sqrt{3}k_{y}\right) \nonumber \\
&  +36t^{2}\cos \left(  3k_{x}/2\right)  \cos \left(  \sqrt{3}k_{y}/2\right)
\cos \left(  \sqrt{3}k_{y}\right) \nonumber \\
&  +2\left(  5\Delta^{2}+27t^{2}\right)  \cos \left(  3k_{x}/2\right)
\cos \left(  \sqrt{3}k_{y}/2\right) \nonumber \\
&  +9t^{2}\cos \left(  3k_{x}\right)  \left(  1+\cos \left(  \sqrt{3}%
k_{y}\right)  \right)  ].
\end{align}
$\chi^{\perp}$ is the transverse spin susceptibility, of which $\zeta$ is
\begin{equation}
\zeta=\frac{1}{N}\sum \limits_{\mathbf{k}}\frac{\Delta^{2}}{4\left(  \left \vert
\xi_{\mathbf{k}}\right \vert ^{2}+\Delta^{2}\right)  ^{\frac{3}{2}}}.
\end{equation}
One can see detailed calculations of $\rho_{s}$ and $\zeta$ in the appendix.

The numerical results of $\rho_{s}$ and $c$ are illustrated in Fig.3, where
one can find that $c=0.168901985t=1.055637J$ in the strongly coupling limit
$\left(  U=25t\right)  $ match the earlier results, $c=1.06066J$
($J=\frac{4t^{2}}{U}$), obtained from Heisenberg model \cite{Mattsson}%
.\begin{figure}[ptb]
\includegraphics[width=0.45\textwidth]{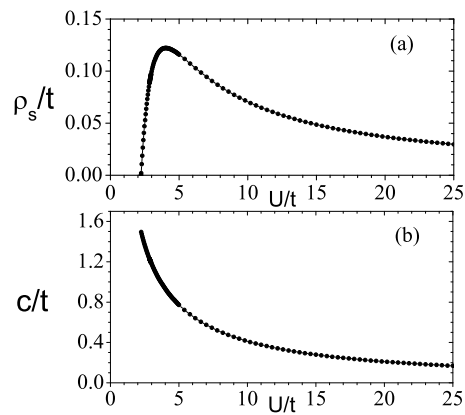}\caption{The spin stiffness
$\rho_{s}$ and the spin wave velocity $c$ of the half-filled honeycomb Hubbard
model at $T=0$.}%
\end{figure}

In addition, we need to determine another important parameter - the cutoff
$\Lambda$. On the one hand, the effective $\mathrm{NL}\sigma \mathrm{M}$ is
valid within the energy scale of Mott gap, $2\Delta=UM.$ On the other hand,
the lattice constant is a natural cutoff. Thus the cutoff is defined as the
following equation \cite{dupuis}%
\begin{equation}
\Lambda=\min(1,\frac{2\Delta}{c})
\end{equation}

\begin{figure}[ptb]
\includegraphics[width=0.45\textwidth]{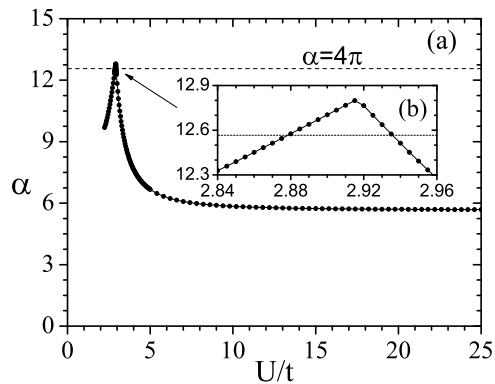}\caption{The dimensionless
coupling constant $\alpha$ of Hubbard model on honeycomb lattice of half
filling at $T=0$. Insert (b) shows non-magnetic region. }%
\end{figure}

\section{Magnetic properties of the insulator state}

In this section we will use the effective $\mathrm{NL}\sigma \mathrm{M}$ to
study the magnetic properties of the insulator state\cite{com1}. The
Lagrangian\ of $\mathrm{NL}\sigma \mathrm{M}$ with a constraint ($\mathbf{n}%
^{2}=1$) by a Lagrange multiplier $\lambda$ becomes
\begin{equation}
\mathcal{L}_{\mathrm{eff}}=\frac{1}{2cg}\left[  \left(  \partial_{\tau
}\mathbf{n}\right)  ^{2}+c^{2}\left(  \mathbf{\bigtriangledown n}\right)
^{2}+i\lambda(1-\mathbf{n}^{2})\right]
\end{equation}
where $i\lambda=m^{2}$ and $m$ is the mass gap of the spin fluctuations.

Using the large-\emph{N} approximation we rescale the field
$\mathbf{n\rightarrow}\sqrt{N}\mathbf{n}$ and obtain the saddle-point equation
of motion as
\begin{equation}
\left(  n_{0}\right)  ^{2}-k_{\mathrm{B}}T\sum_{\omega_{n},\mathbf{q\neq0}}%
\Pi(\mathbf{q},i\omega_{n})=1. \label{eta}%
\end{equation}
In Eq.(\ref{eta}), $n_{0}$ is the mean field value of $\mathbf{n}$ and
$\Pi(\mathbf{q},i\omega_{n})=-\frac{gc}{\omega_{n}^{2}+c^{2}\mathbf{q}%
^{2}+m^{2}}$ is the propagator of the spin fluctuations $\delta \mathbf{n=n-}%
n_{0}.$ Here $\omega_{n}=2\pi nk_{\mathrm{B}}T$, $n=$
integers.\begin{figure}[ptb]
\includegraphics[width=0.45\textwidth]{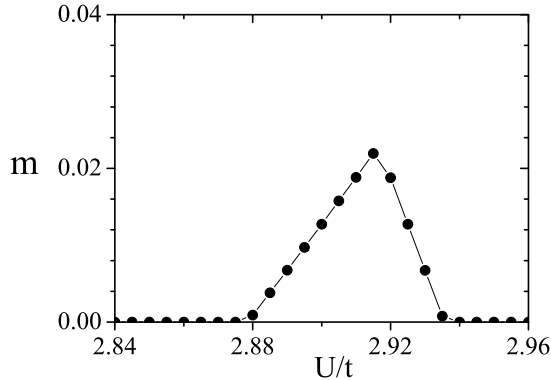}\caption{The mass gap $m$ of
the spin fluctuations of Hubbard model on honeycomb lattice of half filling at
$T=0.$ }%
\end{figure}

At finite temperature, the solution of $n_{0}$ is always zero that is
consistent to the Mermin-Wigner theorem. From Eq.(\ref{eta}), we may get the
solution of $m$ as
\begin{equation}
m=2k_{\mathrm{B}}T\sinh^{-1}\left[  e^{-\frac{2\pi c}{gk_{\mathrm{B}}T}}%
\sinh \left(  \frac{c\Lambda}{2k_{\mathrm{B}}T}\right)  \right]  . \label{mas}%
\end{equation}
In the limit $T\ll \Lambda,$ Eq.(\ref{mas}) can be rewrite as:
\begin{equation}
m=2k_{\mathrm{B}}T\sinh^{-1}\left \{  \frac{1}{2}\exp \left[  -\frac{2\pi
c}{k_{\mathrm{B}}T}\left(  \frac{1}{g}-\frac{1}{g_{c}}\right)  \right]
\right \}
\end{equation}
where
\begin{equation}
g_{c}=\frac{4\pi}{\Lambda}.
\end{equation}
Therefore, at zero temperature the solutions of $n_{0}$ and $m$ of
Eq.(\ref{eta}) are determined by the dimensionless coupling constant
$\alpha=g\Lambda.$ In particular, there exists a critical point $\alpha
_{c}=4\pi$ (or $g_{c}=\frac{4\pi}{\Lambda}$): For the case of $\alpha<4\pi,$
we get solutions of $n_{0}$ and $m$:%
\begin{equation}
n_{0}=(1-\frac{g}{g_{c}})^{1/2}\text{, }m=0
\end{equation}
For the case of $\alpha>4\pi,$ we get solutions of $n_{0}$ and $m$:%
\begin{equation}
n_{0}=0\text{, }m=4\pi c(\frac{1}{g_{c}}-\frac{1}{g})
\end{equation}
\begin{figure}[ptb]
\includegraphics[width=0.45\textwidth]{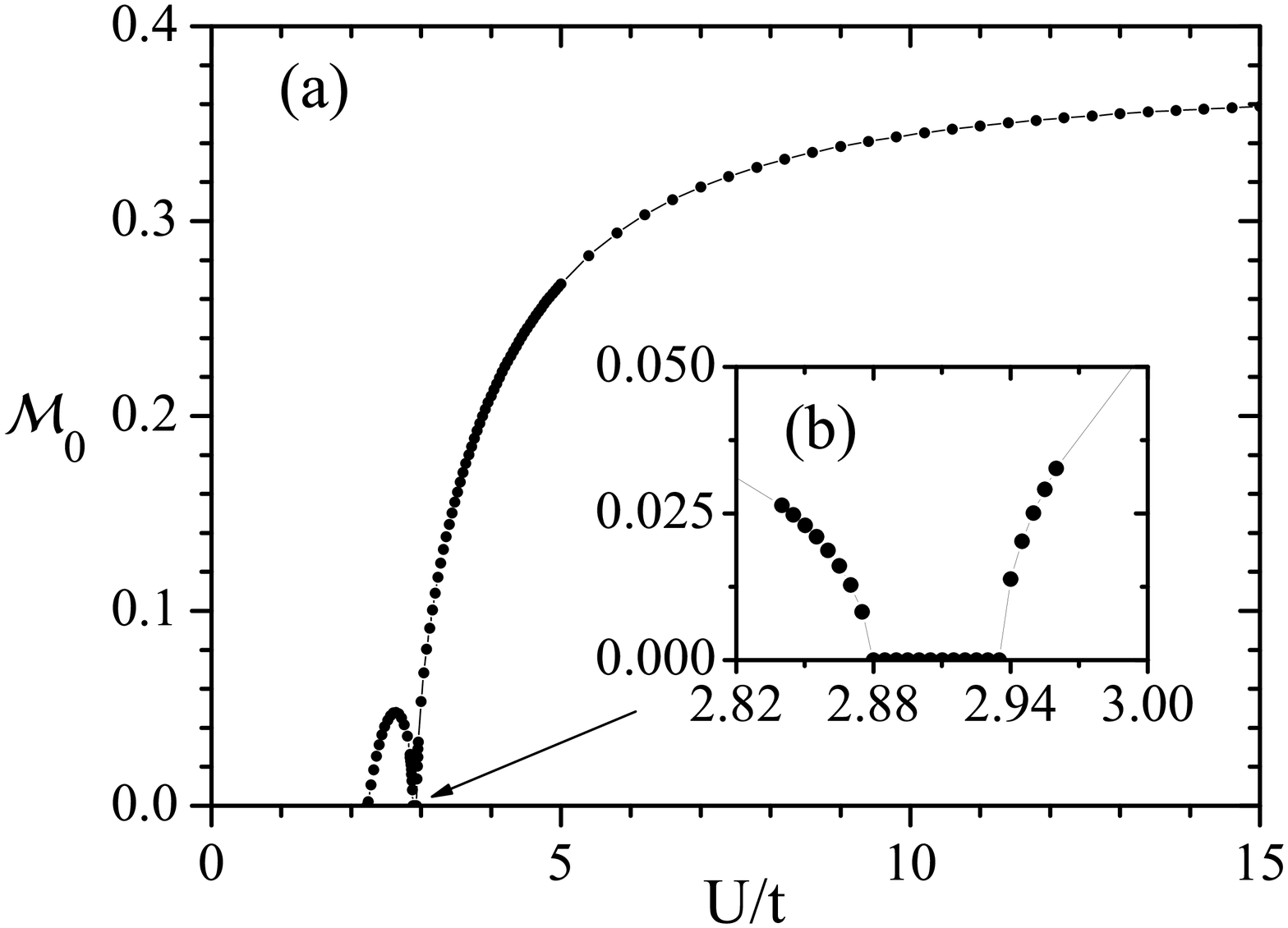}\caption{Spin order parameter
$\mathcal{M}_{0}$ of the Hubbard model on honeycomb lattice of half filling at
$T=0.$ Insert (b) shows non-magnetic region. }%
\end{figure}

So we calculate the dimensionless coupling constant $\alpha=g\Lambda$ and show
results in Fig.4. The quantum critical points corresponding to $\alpha
_{c}=4\pi$ turn into $\left(  U/t\right)  _{c2}\simeq2.88$ and $\left(
U/t\right)  _{c3}\simeq2.93$ which divide the insulator state into three
phases - a quantum disordered state (QD) in the region of $2.88<U/t<2.93$ and
two long range AF order in the regions of $2.23<U/t<2.88$ and $U/t>2.93$.

In the regions of $2.23<U/t<2.88$ and $U/t>2.93$ (where $\alpha<\alpha_{c}$),
at low temperature the mass gap $m$ of spin fluctuations is determined by:
\begin{equation}
m\simeq2k_{\mathrm{B}}T\exp \left[  -\frac{2\pi c}{k_{\mathrm{B}}T}\left(
\frac{1}{g}-\frac{1}{g_{c}}\right)  \right]
\end{equation}
\ Because the energy scale of the mass gap $m$ is always much smaller than the
temperature, \textit{i.e.}, $m\ll k_{\mathrm{B}}T$ (or $\omega_{n}$), quantum
fluctuations become negligible in a sufficiently long wavelength and low
energy regime $\left(  m<\left \vert c\mathbf{q}\right \vert <k_{\mathrm{B}%
}T\right)  .$ Thus in this region one may only consider the purely static
(semiclassical) fluctuations.

At zero temperature, the mass gap $m$ vanishes (See Fig.5), which means that
long range AF order appears. To describe the AF order, we introduce a spin
order parameter\cite{cha,Chubukov,sech}:
\begin{equation}
\mathcal{M}_{0}=\frac{M}{2}n_{0}=\frac{M}{2}(1-\frac{g}{g_{c}})^{1/2}%
\end{equation}
As shown in Fig.6, the ground state of AF ordered phase has a finite spin
order parameter.

In the region of $2.88<U/t<2.93$ (where $\alpha>\alpha_{c}$), there is a
finite mass gap of spin fluctuations at zero temperature (See Fig.5):%
\begin{equation}
m=4\pi c(\frac{1}{g_{c}}-\frac{1}{g})
\end{equation}
Therefore, the ground state of the insulator in this region is not a long
range AF order. Instead, it is a quantum disordered state (or non-magnetic
insulator state) with zero spin order parameter $\mathcal{M}_{0}=0$ in a
\emph{narrow non-magnetic window }(See Fig.6).

Based on above results, we get the global phase diagram which is illustrated
in Fig.7. One can see that at finite T, there are four crossover lines,
$T_{HF}$, $T_{\rho}$, $T_{\nu}$, that separate five regions.

\begin{figure}[ptb]
\includegraphics[width=0.45\textwidth]{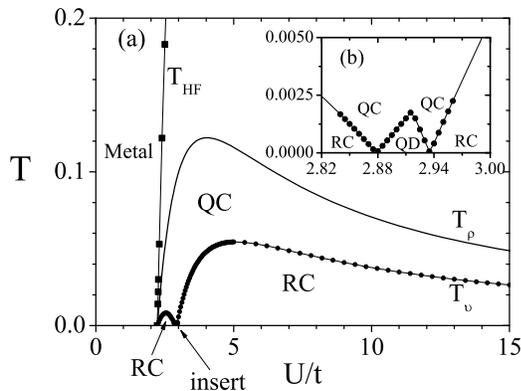}\caption{Phase diagram of the
Hubbard model on honeycomb lattice of half filling at finite temperature.
Insert (b) shows non-magnetic region. }%
\end{figure}

The highest crossover line is $T_{HF}$ that is obtained from Eq.(\ref{self})
and denotes the establish of the effective spin-moments. Above $T_{HF},$ it is
metal phase without energy gap $\Delta=0.$ The crossover line $T_{\rho}%
\sim \rho_{s}$ denotes the validity of the $\mathrm{NL}\sigma \mathrm{M,}$ where
$\rho_{s}$ is the energy scale of spin stiffness. In the region $T_{\rho
}<T<T_{HF},$ the free spin-moments are established (denoted by $M\neq0)$ that
show a Curie-Weiss behavior. In this region one cannot use the effective
$\mathrm{NL}\sigma \mathrm{M}$. Below $T_{\rho},$ short range spin-correlation
exists and the effective $\mathrm{NL}\sigma \mathrm{M}$ is valid. The region
below $T_{\rho}$ is dominated by the crossover lines $T_{\nu}\sim \rho
_{s}\left \vert 1-g/g_{c}\right \vert $ together with two quantum critical
points (QCP) at $\left(  U/t\right)  _{c2}=2.88$ and $\left(  U/t\right)
_{c3}=2.93$ which represent the QCP of $g=g_{c}$\ in the $\mathrm{NL}%
\sigma \mathrm{M.}$\cite{cha,sech}

\section{Spin-spin correlations}

In order to make our theoretical predictions verifiable, we discuss the
detection of the anomalous spin dynamics in ultracold atom experiments via
spatial spin-spin correlations $\left \langle \mathbf{S}\left(  \mathbf{r}%
,t\right)  \cdot \mathbf{S}\left(  0,0\right)  \right \rangle =\left \langle
c^{\dagger}\left(  \mathbf{r},t\right)  \mathbf{\sigma}c\left(  \mathbf{r}%
,t\right)  \cdot c^{\dagger}\left(  0,0\right)  \mathbf{\sigma}c\left(
0,0\right)  \right \rangle $ or dynamic spin susceptibility $\chi^{\prime
\prime}\left(  \mathbf{q},\omega \right)  $. Here $\chi^{\prime \prime}\left(
\mathbf{q},\omega \right)  $ is defined as
\begin{equation}
\chi^{\prime \prime}\left(  \mathbf{q},\omega \right)  =\frac{1}{2}\left(
1-e^{-\omega \beta}\right)  \int dtd\mathbf{r}e^{i\left(  \omega
t-\mathbf{q\cdot r}\right)  }\left \langle \mathbf{S}\left(  \mathbf{r}%
,t\right)  \cdot \mathbf{S}\left(  0,0\right)  \right \rangle .
\label{spin correlation}%
\end{equation}

On the one hand, people may observe spatial spin-spin correlations
$\langle \hat{S}_{z}\left(  \mathbf{r}_{1}\right)  \hat{S}_{z}\left(
\mathbf{r}_{2}\right)  \rangle$ by noise correlation
spectroscopy\cite{noise1,noise2}. As demonstrated in Ref. \cite{noise1}, using
a probe laser beam which goes through the system, one can measure the phase
shift or change of polarization of the outgoing beam $\langle \hat
{X}_{\text{out}}\rangle \propto \langle \hat{M}_{z}\rangle$ to obtain the
magnetization $\langle \hat{M}_{z}\rangle \propto \int d\mathbf{r}\phi \left(
\mathbf{r}\right)  \langle \hat{S}_{z}\left(  \mathbf{r}\right)  \rangle$,
where $\phi \left(  \mathbf{r}\right)  $ is the spatial intensity profile of
the laser beam. The quantum noise $\langle \hat{X}_{\text{out}}^{2}%
\rangle \propto \langle \hat{M}_{z}^{2}\rangle \propto \int d\mathbf{r}%
_{1}d\mathbf{r}_{2}\phi \left(  \mathbf{r}_{1}\right)  \phi \left(
\mathbf{r}_{2}\right)  \langle \hat{S}_{z}\left(  \mathbf{r}_{1}\right)
\rangle \langle \hat{S}_{z}\left(  \mathbf{r}_{2}\right)  \rangle$ reveals the
atomic correlations in the system. This method can directly examine the
existence of 2D AF correlations.

\begin{figure}[ptb]
\includegraphics[width=0.45\textwidth]{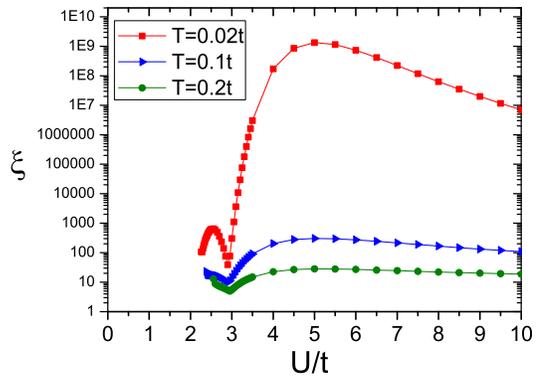}\caption{Correlation lengh
$\xi$ at T = 0.02, 0.1, 0.2. The unit of temperature is $t$.}%
\end{figure}

On the other hand, people may also observe dynamic spin susceptibility
$\chi^{\prime \prime}\left(  \mathbf{q},\omega \right)  $ which links with
spin-spin correlations by NMR measurement. Consequently, it determines the
spin-spin correlation in the real space as $\cos \left(  \mathbf{Q}_{0}%
\cdot \mathbf{r}\right)  e^{-\left \vert \mathbf{r}\right \vert /\xi}$, with the
correlation length $\xi$ and the AF wave-vector $\mathbf{Q}_{0}=\left(
\pm \frac{\pi}{a},\pm \frac{\pi}{a}\right)  $. To show clearly the anomalous
spin dynamics, we calculate the spin-spin correlation $\left \langle
\mathbf{S}\left(  \mathbf{r},t\right)  \cdot \mathbf{S}\left(  0,0\right)
\right \rangle $. The dynamic spin susceptibility at $\mathbf{Q}_{0}$ is
determined by $n$ fields
\begin{equation}
\chi(\mathbf{q},\omega)=\langle n_{a}(\mathbf{q},\omega)n_{a}(-\mathbf{q}%
,-\omega)\rangle=\frac{gc}{\mathbf{q}^{2}+\omega^{2}+m^{2}}%
\end{equation}
where we have introduced the momentum $\mathbf{q}$. The equal-time spin-spin
correlation function is proportion to $e^{-\left \vert \mathbf{r}\right \vert
/\xi}$ where the spin-correlation length $\xi$ is therefore given by $\frac
{1}{m}.$ As Eq.(\ref{spin correlation}) indicates, the spin dynamic structure
factor as well as the dynamic spin susceptibility function may reflect the
effective short-range magnetic correlation length $\xi$.

In Fig.8, we give the spin-correlation length defined as $\xi=\frac{1}{m}$ at
$T=$ $0.02t$, $0.1t$, $0.2t$, respectively. From the Fig.8, taking $T=0.02t$
as an example, one can see that the spin-correlation length increases quickly
with increasing\textbf{ }interaction $U/t$. However, the spin-correlation
length doesn't increase monotonously with $U/t$ - it will decrease and reach a
minimum value near the MI transition $U/t\sim \left(  U/t\right)  _{c2}%
\sim \left(  U/t\right)  _{c3}.$ The dip of the spin-correlation length will
indicate the existence of the non-magnetic state near MI transition. When one
increases the interaction further, the spin-correlation increases again and
finally decreases in the strongly interacting limit due to $J\rightarrow0$.
For other cases with higher temperature, $T=$ $0.1t$, $0.2t$, there exist
similar dip structure of the spin-correlation length via $U/t$ which means
that people may observe the anomalous spin dynamics more easily in experiments.

\section{Doping effect}

In this section we will leave from half-filling and study the hole doping
cases. In case of hole-concentration $d=1-n$, the chemical potential $\mu$\ is
not $\frac{U}{2}$ any more. The Hamiltonian becomes
\begin{align}
\mathcal{H}  &  =-\sum \limits_{\left \langle ij\right \rangle }(t_{i,j}%
c_{i}^{\ast}c_{j}+h.c.)-\sum_{i}\Delta_{i}c_{i}^{\ast}\mathbf{\Omega}%
_{i}\mathbf{\cdot \sigma}c_{i}\nonumber \\
&  +\sum_{i}\left(  \frac{Un}{2}-\mu \right)  c_{i}^{\ast}c_{i}.
\end{align}
At this case, we could obtain $M,$ $\mu,$ $\zeta,$ $\rho_{s}$ similarly with
that of half-filling case in $T=0$ as follows,%
\begin{align}
1  &  =\frac{1}{N}\sum \limits_{E_{\mathbf{k}}<\mu}\frac{U}{2E_{\mathbf{k}}%
},\text{ }1-d=\sum \limits_{E_{\mathbf{k}}<\mu}\mathbf{1,}\\
\zeta &  =\frac{1}{N}\sum \limits_{E_{\mathbf{k}}<\mu}\frac{\Delta^{2}%
}{4\left(  \left \vert \xi_{\mathbf{k}}\right \vert ^{2}+\Delta^{2}\right)
^{\frac{3}{2}}},\nonumber \\
\rho_{s}  &  =\frac{1}{N}\sum \limits_{E_{\mathbf{k}}<\mu}\frac{\epsilon^{2}%
}{4(\left \vert \xi_{\mathbf{k}}\right \vert ^{2}+\Delta^{2})^{\frac{3}{2}}%
}.\nonumber
\end{align}
\begin{figure}[ptb]
\includegraphics[width=0.45\textwidth]{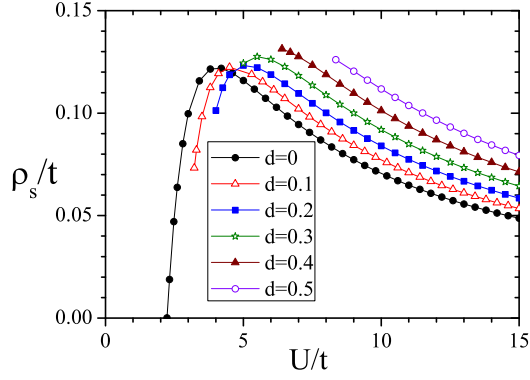}\caption{Spin stiffness
$\rho_{s}$\ of the Hubbard model on honeycomb lattice with hole-concentration
$d=$ 0.0, 0.1, 0.2, 0.3, 0.4, 0.5 at $T=0.$}%
\end{figure}\begin{figure}[ptb]
\includegraphics[width=0.45\textwidth]{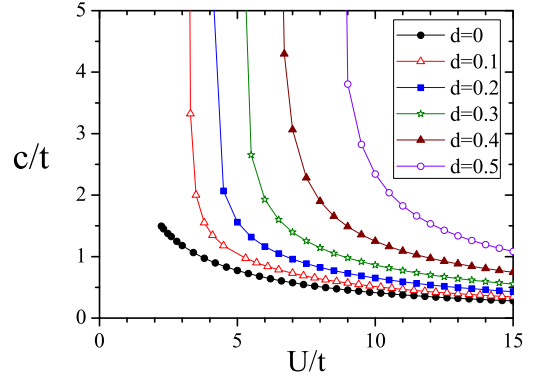}\caption{Spin wave velocity
$c$ of the honeycomb lattice Hubbard model\ with hole-concentration $d=$ 0.0,
0.1, 0.2, 0.3, 0.4, 0.5 at $T=0.$}%
\end{figure}

Given certain hole-concentration $d$, the $\rho_{s}$ and $c$ may be obtained.
From comparation in Fig.9, one can see that spin stiffness $\rho_{s}$ rises
when the hole-concentration increases from $d=0.1$ to $d=0.5,$ and it shows a
different behavior in the large doping density from that of low doping
density. From Fig.10, the spin wave velocity $c$ goes up as well when we
increase hole concentration and it becomes much larger at critical points. The
dimensionless coupling constant $\alpha$ are calculated in Fig.11 in different
hole-concentrations, from which one can see that all critical points go up as
we increase doping concentration $d$. And more importantly, we find that the
non-magnetic regions widen when the hole-concentration increases.

\begin{figure}[ptb]
\includegraphics[width=0.45\textwidth]{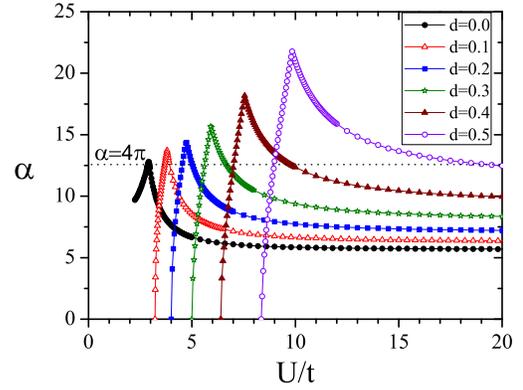}\caption{The dimensionless
coupling constant $\alpha$ of the Hubbard model on honeycomb lattice with
hole-concentration $d=$ 0.0, 0.1, 0.2, 0.3, 0.4, 0.5 at $T=0.$\ }%
\end{figure}

In order to show the doping effect clearly, we plot the phase diagram of the
honeycomb lattice Hubbard model at $T=0$ (See Fig.12). From the phase diagram,
one can see that there are three critical lines ($U_{c1},$ $U_{c2},$ $U_{c3}$)
separating four regimes - one SM regime, two AF regimes and one non-magnetic
regime. In Fig.12, $U_{c1}$ that has been given in Ref.\cite{Peres} is derived
by Hartree-Fock approximation. $U_{c2}$ and $U_{c3}$ are determined by the
critical point of the effective \textrm{NL}$\sigma$\textrm{M}. In particular,
the region of quantum disordered state becomes much wider by increasing the hole-concentration.

\begin{figure}[ptb]
\includegraphics[width=0.45\textwidth]{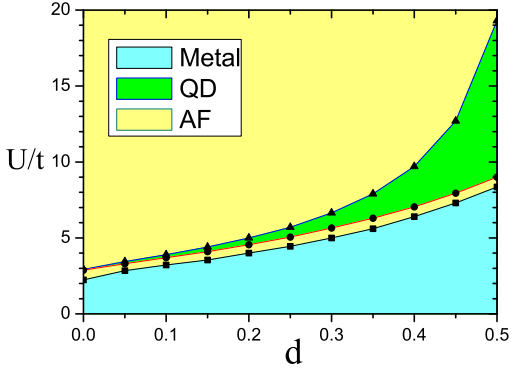}\caption{Phase diagram of the
Hubbard model on honeycomb lattice\ at different hole-concentration at $T=0.$
The square dot line, circle dot line, triangle dot line represent
$U_{c1},U_{c2},U_{c3}$, respectively.}%
\end{figure}

\section{Conclusion}

In this paper, we investigate the two dimensional honeycomb Hubbard model with
an approach that keeps spin rotation symmetry. By it, we found anomalous spin
dynamics not far from the critical point of MI transition : there\ may exist a
narrow non-magnetic insulator. The narrow non-magnetic insulator will be in
favor of hole-doping. Also we discuss how to observe the anomalous spin
dynamics in cold atoms.

In the end we discuss the nature of the non-magnetic insulator in the two
dimensional honeycomb Hubbard model. In Ref.\cite{kou1}, it is pointed out
that by considering the fermionic nature of vortices (half-skyrmions), nodal
spin liquid becomes the ground state of the non-magnetic insulator state in
the $\pi$-flux Hubbard model on square lattice. There exist three types of
quasi-particles in nodal spin liquids: nodal fermionic spinons, gapped bosonic
spinons and roton-like $U(1)$ gauge field. Following the similar approach, one
may draw the same conclusion that the non-magnetic insulator in the two
dimensional honeycomb Hubbard model is another example nodal spin liquid with
similar quasi-particles - nodal fermionic spinons, gapped bosonic spinons and
roton-like $U(1)$ gauge field.

\begin{acknowledgments}
The authors acknowledge stimulating discussions with Z. Y. Weng, T. Li, N. H.
Tong, F. Yang, P. Ye. This research is supported by NCET, NFSC Grant no.
10574014, 10874017.
\end{acknowledgments}

\begin{appendix}
\section{The detailed calculations of $\rho_{s}$ and $\varsigma$}
To give $\rho_{s}$ and $\varsigma$ for calculation, we choose $U_{i}$ to be%
\begin{equation}
U_{i}=\left(
\begin{array}
[c]{cc}%
z_{i\uparrow}^{\ast} & z_{i\downarrow}^{\ast}\\
-z_{i\downarrow} & z_{i\uparrow}%
\end{array}
\right)  ,
\end{equation}
where $\mathbf{n}_{i}=\mathbf{\bar{z}}_{i}\mathbf{\sigma z}_{i},$
$\mathbf{z}_{i}=\left(  z_{i\uparrow},z_{i\downarrow}\right)  ^{T},$
$\mathbf{\bar{z}}_{i}\mathbf{z}_{i}\mathbf{=1.}$ And the spin fluctuations
around $\mathbf{n}_{i}=\mathbf{\hat{z}}_{i}$ is
\begin{align}
\mathbf{n}_{i} &  =\mathbf{\hat{z}}_{i}\mathbf{+}\text{Re}\left(
\mathbf{\phi}_{i}\right)  \mathbf{\hat{x}+}\text{Im}\left(  \mathbf{\phi}%
_{i}\right)  \mathbf{\hat{y}}\\
\mathbf{z}_{i} &  =\left(
\begin{array}
[c]{c}%
1-\left \vert \mathbf{\phi}_{i}\right \vert ^{2}/8\\
\mathbf{\phi}_{i}/2
\end{array}
\right)  +O\left(  \mathbf{\phi}_{i}^{3}\right)  .
\end{align}
Then the quantities $U_{i}^{\dagger}U_{j}$ and $U_{i}^{\dagger}\partial_{\tau
}U_{i}$ can be expanded in the power of $\mathbf{\phi}_{i}-\mathbf{\phi}_{j}$
and $\partial_{\tau}\mathbf{\phi}_{i},$%
\begin{align}
U_{i}^{\dagger}U_{j} &  =e^{i\frac{\mathbf{\phi}_{i}-\mathbf{\phi}_{j}}%
{2}\sigma_{y}}\\
U_{i}^{\dagger}\partial_{\tau}U_{i} &  =\left(
\begin{array}
[c]{cc}%
0 & -\frac{1}{2}\partial_{\tau}\mathbf{\phi}_{i}\\
\frac{1}{2}\partial_{\tau}\mathbf{\phi}_{i} & 0
\end{array}
\right)  .
\end{align}
According to Eq.(\ref{ao}), the gauge field $a_{ij}$ and $a_{0}\left(
i\right)  $ are given as
\begin{align}
a_{ij} &  =\frac{1}{2}\left(  \mathbf{\phi}_{i}-\mathbf{\phi}_{j}\right)
\mathbf{\sigma}_{y}\\
a_{0}\left(  i\right)   &  =\frac{i}{2}\partial_{\tau}\mathbf{\phi}%
_{i}\mathbf{\sigma}_{y}.
\end{align}
Supposing $a_{ij}$ and $a_{0}\left(  i\right)  $ to be a constant in space and
denoting $\mathbf{\partial}_{i}\mathbf{\mathbf{\phi}}_{i}\mathbf{=a}$ and
$\partial_{\tau}\mathbf{\phi}_{i}=iB_{y}$, we have
\begin{align}
a_{ij} &  =-\frac{1}{2}\mathbf{a\cdot(i-j)\sigma}_{y}\\
a_{0}\left(  i\right)   &  =-\frac{1}{2}B_{y}\mathbf{\sigma}_{y}.
\end{align}
The energy of Hamiltonian of Eq.(\ref{efff}) becomes
\begin{equation}
E\left(  B_{y},\mathbf{a}\right)  =\frac{1}{2}\zeta B_{y}^{2}+\frac{1}{2}%
\rho_{s}\mathbf{a}^{2}.\label{ener}%
\end{equation}
Then one could get $\zeta$ and $\rho_{s}$ from the following equations by
calculating the partial derivative of the energy
\begin{align}
\zeta &  =\frac{1}{N}\frac{\partial^{2}E_{0}\left(  B_{y}\right)  }{\partial
B_{y}^{2}}|_{B_{y}=0}\label{weca}\\
\rho_{s} &  =\frac{1}{N}\frac{\partial^{2}E_{0}\left(  \mathbf{a}\right)
}{\partial \mathbf{a}^{2}}|_{\mathbf{a}=0}.\label{wero}%
\end{align}
Here $E_{0}\left(  B_{y}\right)  $ and $E_{0}\left(  \mathbf{a}\right)  $ are
the energy of the lower Hubbard band%
\begin{align}
E_{0}\left(  B_{y}\right)   &  =\sum \limits_{\mathbf{k}}\left(
E_{+,\mathbf{k}}^{\zeta}+E_{-,\mathbf{k}}^{\zeta}\right)  \label{ek}\\
E_{0}\left(  \mathbf{a}\right)   &  =\sum \limits_{\mathbf{k}}\left(
E_{+,\mathbf{k}}^{\rho}+E_{-,\mathbf{k}}^{\rho}\right)  \label{er}%
\end{align}
where $E_{+,\mathbf{k}}^{\zeta},$ $E_{-,\mathbf{k}}^{\zeta}$ and
$E_{+,\mathbf{k}}^{\rho},$ $E_{-,\mathbf{k}}^{\rho}$ are the energies of the
following Hamiltonian $\mathcal{H}^{\zeta}$ and $\mathcal{H}^{\rho}$%
\begin{align}
\mathcal{H}^{\zeta}  &  =-\sum \limits_{<ij>}\left(  t_{i,j}\psi_{i}^{\ast}%
\psi_{j}+h.c.\right)  -\Delta \sum \limits_{i}(-1)^{i}\psi_{i}^{\ast
}\mathbf{\sigma}_{z}\psi_{i}\nonumber \\
&  +\sum \limits_{i}\psi_{i}^{\ast}a_{0}\left(  i\right)  \psi_{i} \label{hca}%
\end{align}%
\begin{equation}
\mathcal{H}^{\rho}=-\sum \limits_{<ij>}\left(  t_{i,j}\psi_{i}^{\ast}e^{a_{ij}%
}\psi_{j}+h.c.\right)  -\Delta \sum \limits_{i}(-1)^{i}\psi_{i}^{\ast
}\mathbf{\sigma}_{z}\psi_{i}. \label{hrou}%
\end{equation}
Using the Fourier transformations for $\mathcal{H}^{\zeta}$ , we have the
spectrum of the lower band of $\mathcal{H}^{\zeta}$%
\begin{equation}
E_{\pm,\mathbf{k}}^{\zeta}=-\sqrt{\left(  \left \vert \xi_{\mathbf{k}%
}\right \vert \pm \frac{B_{y}}{2}\right)  ^{2}+\Delta^{2}}%
\end{equation}
where $\xi_{\mathbf{k}}$ has been obtained in Eq.(\ref{kesi}). Using
Eq.(\ref{weca}), $\zeta$ is obtained as
\begin{equation}
\zeta=\frac{1}{N}\sum \limits_{\mathbf{k}}\frac{\Delta^{2}}{4\left(  \left \vert
\xi_{\mathbf{k}}\right \vert ^{2}+\Delta^{2}\right)  ^{\frac{3}{2}}}.
\label{pic}%
\end{equation}
Similarly, using the Fourier transformations for $\mathcal{H}^{\rho},$ we
obtain the spectrum of the lower band of $\mathcal{H}^{\rho}$
\begin{equation}
E_{\pm,\mathbf{k}}^{\rho}=-\sqrt{\Delta^{2}+\left \vert \psi \right \vert
^{2}+\left \vert \varphi \right \vert ^{2}\pm \left[  4\Delta^{2}\left \vert
\psi \right \vert ^{2}-\left(  \varphi \psi^{\ast}-\psi \varphi^{\ast}\right)
^{2}\right]  ^{\frac{1}{2}}}%
\end{equation}
where $\varphi$ and $\psi$ are defined as
\begin{align}
\varphi &  =-t\sum \limits_{\mathbf{\delta}}e^{i\mathbf{k\cdot \delta}}%
\cos \left(  \frac{1}{2}\mathbf{a\cdot \delta}\right) \\
\psi &  =-t\sum \limits_{\mathbf{\delta}}e^{i\mathbf{k\cdot \delta}}\sin \left(
\frac{1}{2}\mathbf{a\cdot \delta}\right)  .
\end{align}
Using Eq.(\ref{wero}), $\rho_{s}$ is given as
\begin{equation}
\rho_{s}=\frac{1}{N}\sum \limits_{\mathbf{k}}\frac{\epsilon^{2}}{4(\left \vert
\xi_{\mathbf{k}}\right \vert ^{2}+\Delta^{2})^{\frac{3}{2}}}. \label{ita}%
\end{equation}
The corresponding coefficient $\epsilon^{2}$ is
\begin{align}
\epsilon^{2}  &  =\frac{1}{4}t^{2}[6\Delta^{2}+27t^{2}+\left(  2\Delta
^{2}+27t^{2}\right)  \cos \left(  \sqrt{3}k_{y}\right) \nonumber \\
&  +36t^{2}\cos \left(  3k_{x}/2\right)  \cos \left(  \sqrt{3}k_{y}/2\right)
\cos \left(  \sqrt{3}k_{y}\right) \nonumber \\
&  +2\left(  5\Delta^{2}+27t^{2}\right)  \cos \left(  3k_{x}/2\right)
\cos \left(  \sqrt{3}k_{y}/2\right) \nonumber \\
&  +9t^{2}\cos \left(  3k_{x}\right)  \left(  1+\cos \left(  \sqrt{3}%
k_{y}\right)  \right)  ].
\end{align}
\end{appendix}


\vskip1.5cm

\begin{thebibliography}{99}                                                                                               %


\bibitem {Bloch}I. Bloch, J. Dalibard and W. Zwerger Rev. Mod. Phys.
\textbf{80} 885 (2008).

\bibitem {review}S. Giorgini, L. P. Pitaevskii, \& S. Stringari, \textit{Rev.
Mod. Phys.} \textbf{80,} 1215 (2008).

\bibitem {Greiner}M. Greiner, O. Mandel, T. Esslinger, T. W. H\"{a}nsch, \& I.
Bloch, \textit{Nature} \textbf{415,} 39-44 (2002).

\bibitem {Esslinger}R. J\"{o}dens, N. Strohmaier, K. G\"{u}nter, H. Moritz, \&
T. A Esslinger, \textit{Nature} \textbf{455,} 204-207 (2008).

\bibitem {Bloch2}U. Schneider, Science \textbf{322}, 1520-1525 (2008).

\bibitem {Iskin}M. Iskin and C. J. Williams, Phys. Rev. A \textbf{78},
011603(R) (2008).

\bibitem {Andersen}B. M. Andersen and G. M. Bruun, Phys. Rev. A \textbf{76},
041602(R) (2007).

\bibitem {Snoek}M. Snoek, I. Titvinidze, C. Toke, K. Byczuk, and W.
Hofstetter, New J. Phys. \textbf{10}, 093008 (2008).

\bibitem {Gottwald}T. Gottwald and P. G. J. van Dongen, Phys. Rev. A
\textbf{80}, 033603 2009.

\bibitem {duan}L.-M. Duan, E. Demler, and M. D. Lukin, Phys. Rev. Lett.
\textbf{91}, 090402 (2003).

\bibitem {Zhu}S. L. Zhu, B. G Wang, and L.-M. Duan, Phys. Rev. Lett.
\textbf{98}, 260402 (2007).

\bibitem {Lee}Kean Loon Lee \emph{et al.}, A \textbf{80}, 043411 (2009).

\bibitem {Hermele}M. Hermele, Phys. Rev. B, \textbf{76} 035125 (2007).

\bibitem {Sorella}S. Sorella and E. Tosatti, Europhys. Lett. \textbf{19}, 699 (1992).

\bibitem {Martelo}L. M. Martelo, \emph{et al.}, Z. Phys. B: Condens. Matter
\textbf{103}, 335 (1997).

\bibitem {Paiva}T. Paiva, \emph{et al.}, Phys. Rev. B \textbf{72}, 085123 (2005).

\bibitem {dup}N. Dupuis, Phys. Rev. B \textbf{65}, 245118 (2002).

\bibitem {dupuis}K. Borejsza, N. Dupuis, Euro Phys. Lett. \textbf{63}, 722
(2003). K. Borejsza and N. Dupuis Phys. Rev. B \textbf{69}, 085119 (2004).

\bibitem {sun}G. Y Sun and S. P Kou, Europhys. Lett. \textbf{87} 67002 (2009).

\bibitem {wen}{X. G. Wen}, \emph{Quantum Field Theory of Many-Body Systems},
(Oxford Univ. Press, Oxford, 2004).

\bibitem {Schulz}H. J. Schulz, in \emph{The hubbard Model}, edited by D.
Baeriswyl(Plenum, New York, 1995).

\bibitem {weng}Z. Y. Weng, C. S. Ting, and T. K. Lee, Phys. Rev. B
\textbf{43}, 3790 (1991).

\bibitem {Peres}N .M .R. Peres, \emph{et al.}, Phys. Rev. B 70, 195122 (2004).

\bibitem {Haldane}F. D. M. Haldane, Phys. Lett. \textbf{93A}, 464(1983).

\bibitem {Auerbach}A. Auerbach, \emph{Interacting Electrons and Quantum
Magnetism} (Springer-Verlag, New York, 1994).

\bibitem {Mattsson}Ann Mattsson and Per Fr\H{o}jdh, Phys. Rev. B \textbf{49}, 3997(1994).

\bibitem {com1}Near the critical point that the mean field value of $M$
vanishes, due to the strong charge fluctuations and the strong amplitude
fluctuations of $M$, $\mathrm{NL}\sigma \mathrm{M}$ cannot be used and our
results cannot be reliable.

\bibitem {cha}S. Chakravarty, \emph{et al.}, Phys. Rev. B \textbf{39}, 2344 (1989).

\bibitem {Chubukov}A.V. Chubukov and D.M. Frenkel, Phys. Rev. B \textbf{46}, 11884(1992).

\bibitem {sech}S. Sachdev, \emph{Quantum Phase Transitions}, (Cambridge
University Press, 1999).

\bibitem {noise1}G. M. Bruun, B. M. Andersen, E. Demler, A. S. S\o ensen,
Phys. Rev. Lett. \textbf{102}, 030401 (2009).

\bibitem {noise2}V. Guarrera, \emph{et al.}$\emph{,}$ Phys. Rev. Lett.
\textbf{100}, 250403 (2008).

\bibitem {kou1}S. P Kou, L. F. Liu, arXiv: 0910.2070.
\end{thebibliography}
\end{document}